# $^{1}$H-NMR Study on the Magnetic Order in the Mixture of Two Spin Gap Systems $(CH_3)_2CHNH_3CuCl_3$ and $(CH_3)_2CHNH_3CuBr_3$

Keishi Kanada[1], Takehiro Saito[1], Akira Oosawa[1], Takayuki Goto[1]
and Takao Suzuki[1,2]

[1]*Department of Physics, Faculty of Science and Technology, Sophia University*
*7-1 Kioicho, Chiyodaku Tokyo 102-8554*

[2]*Advanced Meson Science Laboratory, RIKEN*
*2-1 Hirosawa, Wako, Saitama 351-0198*

The antiferromagnetic ordering in the solid-solution of the two spin-gap systems $(CH_3)_2CHNH_3CuCl_3$ and $(CH_3)_2CHNH_3CuBr_3$ has been investigated by $^{1}$H-NMR. The sample with the Cl-content ratio $x$=0.85 showed a clear splitting in spectra below $T_N$=13.5 K, where the spin-lattice relaxation rate $T_1^{-1}$ showed a diverging behavior. The critical exponent of the temperature dependence of the hyperfine field is found to be 0.33.

## 1. Introduction

In classical spin systems, disorder destroys magnetic order by cutting off the correlation between neighboring spins. However, in quantum spin systems, where the magnetic order is instabilized by the quantum spin fluctuation, disorder often brings the non-trivial effect to their ground state. A low number of holes drastically destruct the antiferromagnetic order in parent phase of high-$T_C$ cuprates through frustration effect.[1) ] In spin liquid systems represented by $CuGeO_3$, disorder induces an unpaired spin among singlet dimer and brings a static magnetic order to the disordered state.[2)]

The title compounds of $(CH_3)_2CHNH_3CuCl_3$ and $(CH_3)_2CHNH_3CuBr_3$ abbreviated as IPA-$CuCl_3$ and IPA-$CuBr_3$ are isostructural spin gap systems with a non-magnetic ground state. Both the mechanism and the magnitude of the gap are quite different in the two systems. In the Cl-system, which was discovered to have a ladder-like magnetic structure based on an inelastic neutron experiment[5)], the neighboring two $S$=1/2 spins on the rung interact ferromagnetically to form effective $S$=1 spins at low temperatures. The interaction among these integer spins is weak and antiferromagnetic, so that the ground state is gapped one as Haldane conjectured[6]. In a Br-system, the interaction between the two neighboring spins is antiferromagnetic, so that they form a singlet dimer state at low temperatures. The spin excitation gaps of these two systems have been reported to be 18 and 98 K based on magnetization experiments[7-9]. The former was re-examined in the neutron experiment and reported to be 13.9 K[5]. The aim of this paper is a microscopic investigation on the ground state of the solid solution of the two spin gap systems.

Manaka *et al.* have been working intensively on the macroscopic measurements in IPA-$Cu(Cl_xBr_{1-x})_3$, and they report that the system becomes gapless only within the limited region $x$=0.44-0.87, where a magnetic order occurs at low temperatures[9,10]. However, a microscopic investigation on the spin state has not been conducted except for our preliminary result by μSR[11]. In this paper, we report the existence of an antiferromagnetic phase transition in IPA-$Cu(Cl_xBr_{1-x})_3$ ($x$=0.85) through the use of $^1$H-NMR spectra and the spin-lattice relaxation rate.

## 2. **Experimental**

Evaporation method was utilized to grow single crystals of IPA-$Cu(Cl_xBr_{1-x})_3$ ($x$=1 and 0.85). An isopropanol solution of isopropylamine hydrochloride, copper(II) chloride dihydrate, isopropylamine hydrobromide, copper (II) bromide was placed in a bowl, which was maintained at 30(±0.1)°C, and in an atmosphere of flowing nitrogen gas during an entire period of crystal growth, approximately two months. A typical size of the obtained crystals was around 3×5×10mm with a rectangular shape as was reported in a previous paper [9]. The content of Cl, $x$=0.85(5) was determined using the ICP method for three tiny fragments chipped off from different points of the crystal. Macroscopic quantities of obtained crystals, the specific heat and the susceptibility were measured by PPMS and MPMS manufactured by Quantum Design Co Ltd. The uniform susceptibility of $x$=0.85 showed a kink at around 12K, below which it steeply decreased for $H\perp$C and slightly increases for $H\perp$A. The specific heat also showed a small but distinct cusp at the same temperature. The sample $x$=1 showed no anomalies in neither of the two quantities. These results are consistent with reports by Manaka [9,10].

$^1$H-NMR measurements were performed using a 4K cryogen-free refrigerator set in a 6T cryogen-free-superconducting magnet. The spectra were measured by recording a spin-echo amplitude simultaneously ramping up or down the magnetic field. The spin-lattice relaxation rate $T_1^{-1}$ was measured by the saturation-recovery method with a pulse train. Relaxation curves were traced until the difference of the nuclear magnetization from its saturation value becomes one percent.

## 3. Results and Discussion

Figure 1 shows the field-swept spectra of the two samples $x$=0.85 and 1.0 at various temperatures under the field around 3T. At high temperatures both the samples show a sharp paramagnetic resonance line. The width around 18K is approximately 200 and 80 Oe for $x$=0.85 and 1.0 respectively. As the temperature decreases, the $x$=1.0 sample maintains a sharp width, and only the $x$=0.85 sample shows a drastic splitting in the peak below $T_N$=13.5 K. This temperature coincides with that for an anomaly in $\chi_0$ and the specific heat. Five distinct peaks overlap each other at the lowest temperature of 4 K.

These multiple peaks in spectra correspond to the inequivalent proton sites exposed to a hyperfine field produced by the antiferromagnetically-ordered spins. Ten inequivalent proton sites are in the unit cell, and these proton sites have different distances from the nearest Cu site, varying from 3.3 to 6.8 Å. They should feel different hyperfine fields from the ordered moment and should contribute to each split peak. Actually, observed peak separations around 300 Oe are comparable with the estimation by the classical dipole-dipole interaction between electronic and nuclear spins. Though we are still in the process of conducting a detailed site assignment, which requires measurements of spectra involving various directions of the field, we conclude that the spin structure is simple antiferromagnetic rather than incommensurate or glass-like.

Recently, Nakamura numerically investigated the disordered bond-alternating spin chain to report that an antiferromagnetic state is more stable than a glass-like state expected in a classical point of view[4]. He pointed out that the emergence of Néel state is totally to due the quantum effect. In our previous μSR report[11], we have pointed out that in the sample with $x$=0.95 which belongs to the gapped region, the antiferromagnetic spin fluctuation is anomalously enhanced at low temperatures. This fact suggests that the magnetic instability is inherently present in the gapped sample.

Figure 2 shows the dependence on the temperature of a peak separation $\Delta H$, which is defined as the distance between the positions at 20% of maxima of the left most and right most peaks, and is considered to be a good measure for the magnitude of the staggered moment and hence of the order parameter. The dependence of $\Delta H$ on temperature is aptly described by the scaling function $\left(1-T/T_{\mathrm{N}}\right)^{\beta}$, where $\beta$ is a critical exponent estimated from data fitting to be 0.33, the value of which is close to the value of 0.327 which is expected 3D-Ising model[16], and is consistent with our report by μSR[11]. The reason for the appearance of dimensionality of the universality class is 3D, simply because the phase transition in 1D-spin systems must be set off by the weak interaction paths that runs three dimensionally in the system.

In previous paper[11], we have explained the appearance of the Ising-type universality class in terms of $D$*, an effective single-ion anisotropy that appears at low temperatures owing to the formation of effective $S$=1 spins in ferromagnetic dimers. However, an existence of effective $S$=1 spins[12-14] are proved only in the pure system of $x$=1.0 but the disordered system $x$=0.85 containing both ferromagnetic and antiferromagnetic dimers. It is, therefore, not self-evident whether or not the disordered system has a universality class identical to the pure system. According to Harris criterion[15], the disorder does not affect the universality class only for those systems with the negative critical exponent of the specific heat $\alpha$, which does not hold in the present case.

Therefore, the universality class of $x$=0.85 sample should be determined carefully through experiments. Generally, in the presence of finite magnetic anisotropy $D$, the universality class of the spin system with Heisenberg interaction $J$ shows a crossover from the isotropic Heisenberg class to Ising-like one as temperature getting closer to $T_N$, at the temperature defined as $\left|\frac{D}{J}t^{-\phi}\right| \sim 1$, where $t$ is the reduced temperature, and $\phi < 1$, the crossover index[16]. The critical exponent is expected to show simultaneously the gradual change from 0.367 of 3D-Heisenberg to 0.327 of 3D-Ising. However, as is clear in Fig. 2, the value of $\beta$, the gradient of the fitting line, shows a further decrease from 0.33 as the fitting temperature region is expanded. We conclude that our previous interpretation of the observed $\beta$ to be the Ising model arisen from $D$* is possibly be misleading. The reason of appearance of $\beta$=0.33 experimentally confirmed by both μSR and NMR is still open question at this stage, and must be resolved in the future both theoretically and experimentally.

Figure 3 shows the dependence of the spin-lattice relaxation rate $T_1^{-1}$ on temperature. In a ordered state, $T_1^{-1}$ was measured on a left most peak. There was no significant difference in $T_1^{-1}$ on any peaks except for the center peak which bears rather a lower relaxation rate. While $T_1^{-1}$ of the gapped sample ($x$=1.0) decreases exponentially as the temperature decreases, reflecting the gapped ground state, the gapless sample ($x$=0.85) shows diverging behavior around $T_{\mathrm{N}}$. This clearly demonstrates the fact that the observed phase transition is a second order one. In the vicinity of the second order phase transition, the fluctuation in the magnetic field shows a critical slow down and enhances NMR-$T_1^{-1}$, a measure for the Fourier component of the Larmor frequency in the fluctuation. The dominant $q$-component of the fluctuation is considered to be located at far apart from zero, that is, possibly, around $q = \pi$. The reason for this is the uniform susceptibility that probes the fluctuation around $q$=0 shows no diverging behavior around $T_{\mathrm{N}}$.

The scaling plot of $T_1^{-1}$ in the temperature region both above and below $T_{\mathrm{N}}$ is shown in Fig. 4, where the dynamical critical exponent was obtained by fitting $\left(1-|T|/T_{\mathrm{N}}\right)^n$ to the observed temperature dependence as $n$=1.0(4) for $T < T_{\mathrm{N}}$ and 0.5(4) for $T > T_{\mathrm{N}}$. The latter value above $T_{\mathrm{N}}$ coincides with that for the classical 3D localized spin system. The universality class belongs to the three dimensional one, for which the same argument as that on β stated above holds.

In conclusion, we investigated $^1$H-NMR on a bond-disordered spin-gap system IPA-Cu($Cl_{0.85}Br_{0.15}$)$_3$. The existence of an antiferromagnetic long-range order was clearly demonstrated by peak splitting in the spectra and by the divergence of $T_1^{-1}$. The critical exponent was obtained from the temperature dependence of hyperfine field to be $\beta$=0.33, which is close to the value expected for the 3D-Ising model.

**Acknowledgment**

We thank Dr. H. Manaka and Prof. T. Ohtsuki for their valuable advice, and Dr. K. Noda at Kuwahara Lab., Sophia University for his assistance with specific heat measurements using PPMS. We also thank Prof. T. Nojima at the Center of Low Temperature Science of Tohoku Univ. for his assistance with magnetization measurements using MPMS. This work was supported by the Kurata Memorial Hitachi Science and Technology Foundation, Saneyoshi Scholarship Foundation and by a Grant-in-Aid for Scientific Research on priority Areas "High Field Spin Science in 100T" (No.451) from the Ministry of Education, Culture, Sports, Science and Technology (MEXT).

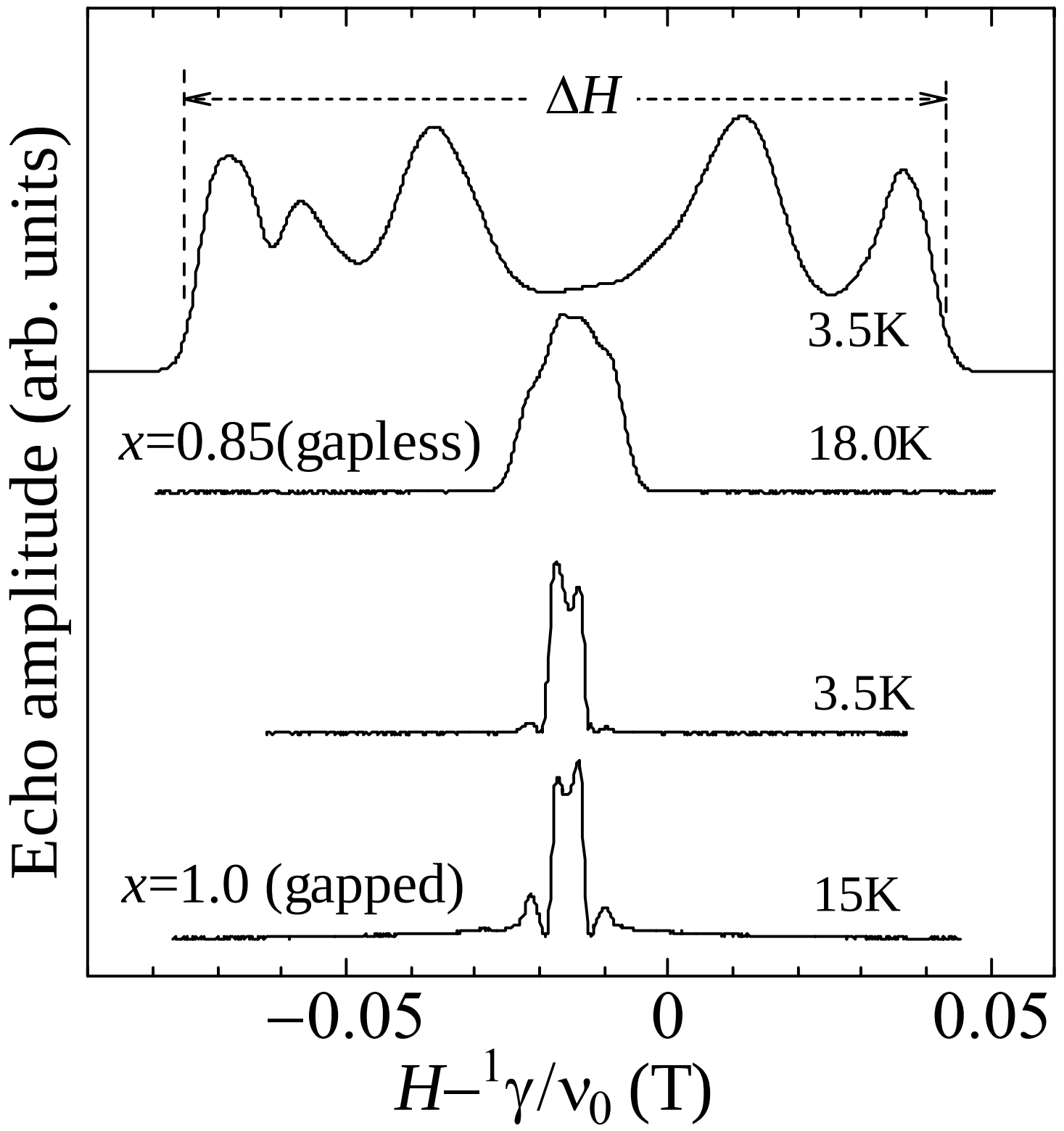

Fig. 1 [1]H-NMR spectra of $x$=0.85 and 1.0 at various temperatures. The gyromagnetic ratio of [1]H, 42.5774MHz/T is denoted as $^1\gamma$.

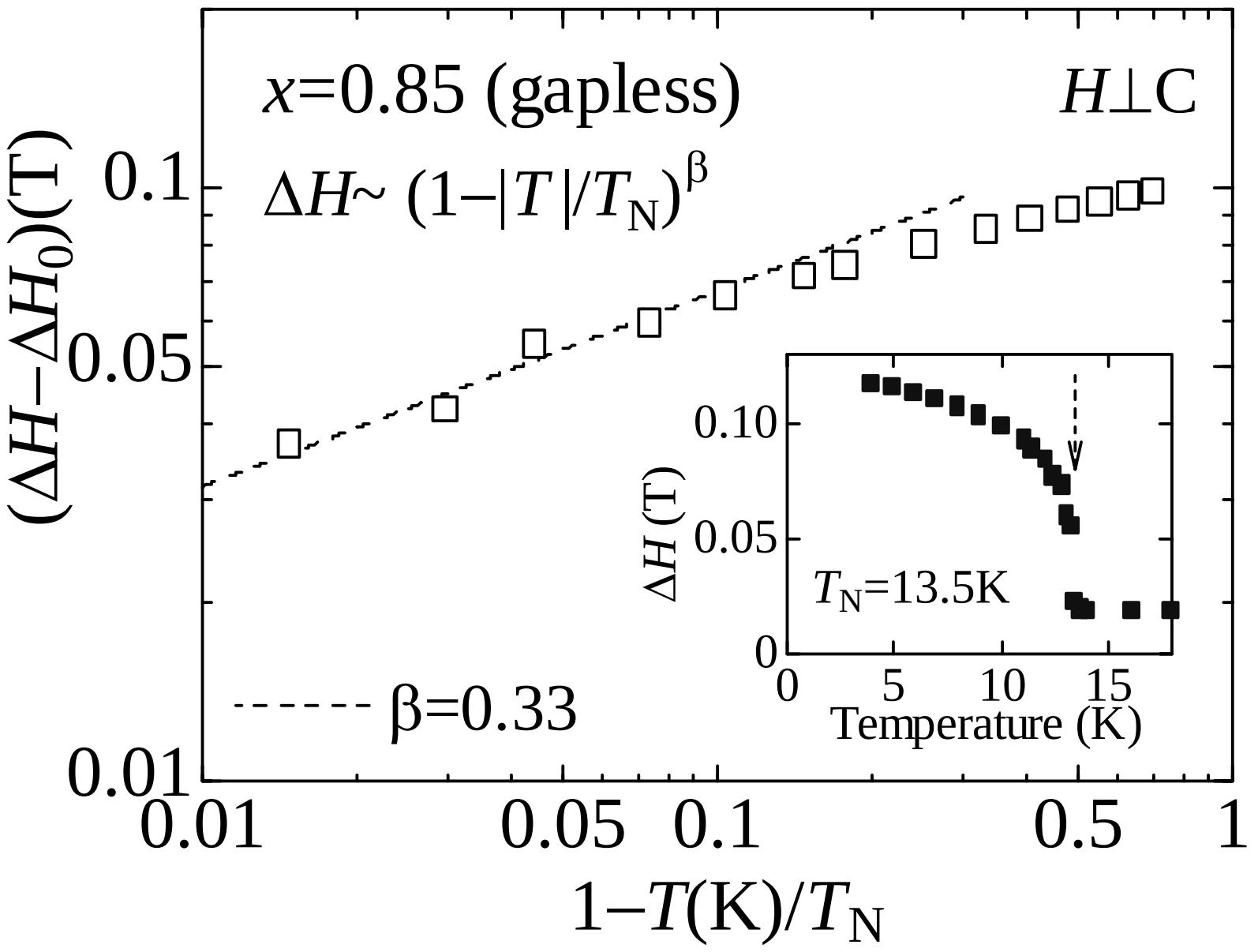

Fig. 2 Dependence of the splitting width $\Delta H$ on the temperature, definition of which is shown in Fig. 1. The curve shows the scaling function $(1 - T/T_N)^\beta$, where $T_N = 13.5$ K and $\beta = 0.33$ are determined by data fitting.

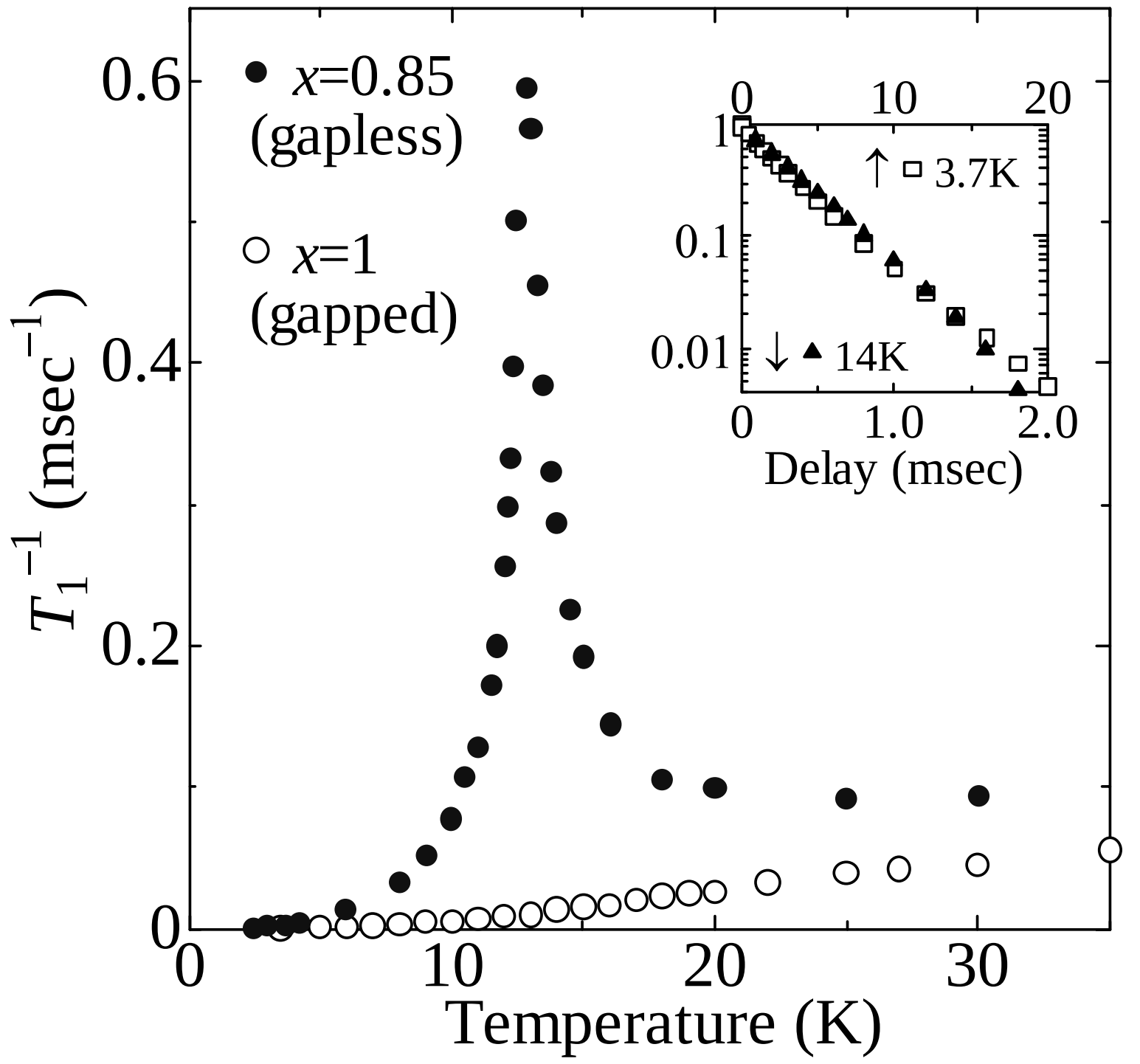

Fig. 3 Dependence of the spin-lattice relaxation rate $T_1^{-1}$ on temperature of gapped ($x$=1.0) and gapless ($x$=0.85) samples with the field direction $H$⊥C and $H$⊥B respectively. The inset shows typical relaxation curves for the $x$=0.85 sample.

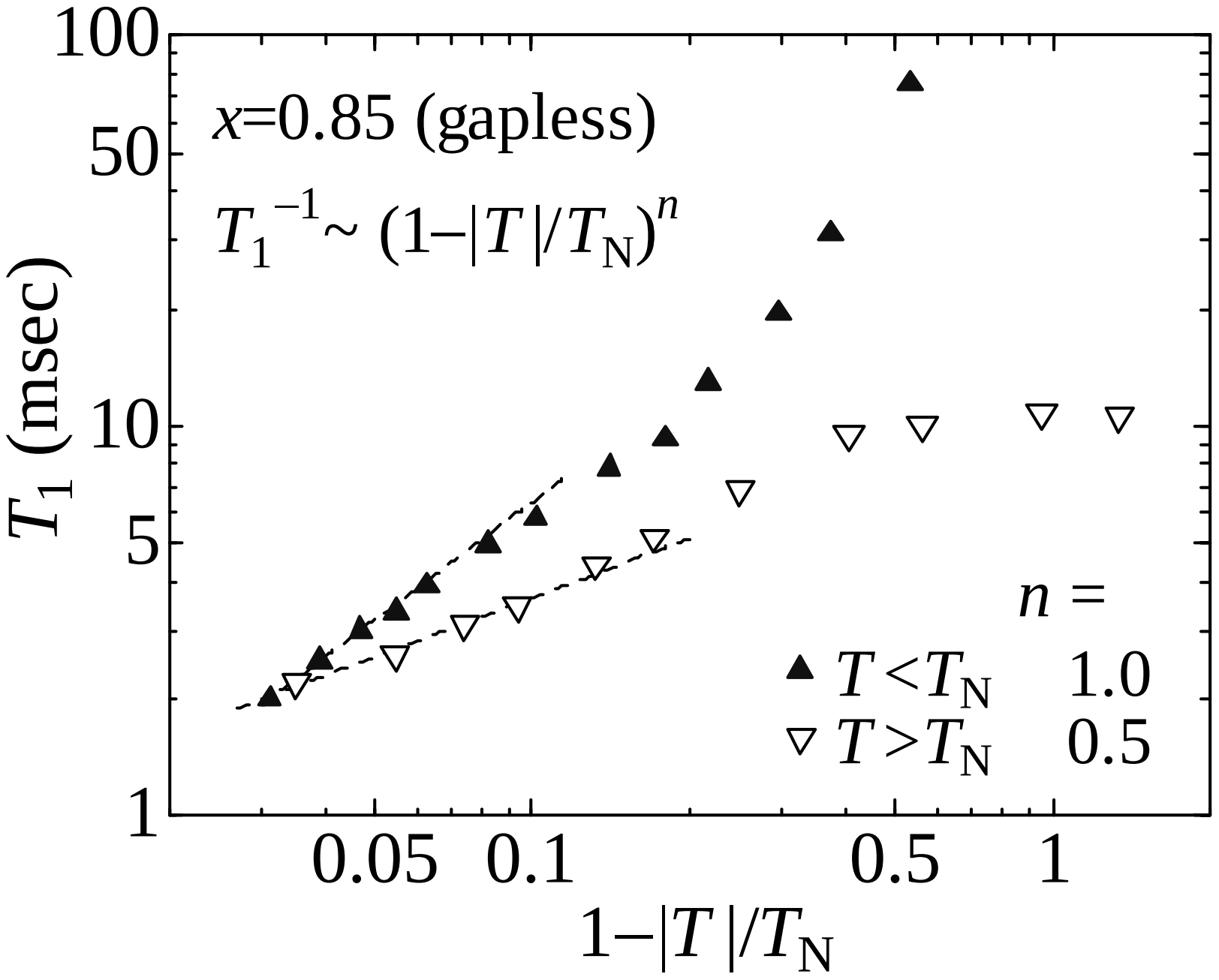

Fig. 4 Scaling plot of $T_1^{-1}$ against the reduced temperature $1-\left|T/T_{\mathrm{N}}\right|$. The dynamical critical exponent was obtained from data fitting (dashed lines).